\documentclass[prb,aps,amsmath,twocolumn,amssymb,floatfixng,showpacs,
superscriptaddress]{revtex4}

\usepackage[dvips]{graphics}
\usepackage{color}
\definecolor{dred}{rgb}{0.5,0,0.5}

\begin{document}

\title{\textcolor{dred}{Persistent charge and spin currents in a quantum 
ring using Green's function technique: Interplay between magnetic flux 
and spin-orbit interactions}}

\author{Santanu K. Maiti}

\email{santanu.maiti@isical.ac.in}

\affiliation{Physics and Applied Mathematics Unit, Indian Statistical
Institute, 203 Barrackpore Trunk Road, Kolkata-700 108, India}

\author{Moumita Dey}

\affiliation{Condensed Matter Physics Division, Saha Institute of Nuclear
Physics, Sector-I, Block-AF, Bidhannagar, Kolkata-700 064, India}

\author{S. N. Karmakar}

\affiliation{Condensed Matter Physics Division, Saha Institute of Nuclear
Physics, Sector-I, Block-AF, Bidhannagar, Kolkata-700 064, India}

\begin{abstract}

We put forward a new approach based on Green's function formalism to 
evaluate precisely persistent charge and spin currents in an Aharonov-Bohm 
ring subjected to Rashba and Dresselhaus spin-orbit interactions. 
Unlike conventional methods our present scheme circumvents direct 
evaluation of eigenvalues and eigenstates of the system Hamiltonian to 
determine persistent currents which essentially reduces possible numerical 
errors, especially for larger rings. The interplay of Aharonov-Bohm flux 
and spin-orbit interactions in persistent charge and spin currents of 
quantum rings is analyzed in detail and our results lead to a possibility 
of estimating the strength of any one of the spin-orbit fields provided 
the other one is known. All these features are exactly invariant even in 
presence of impurities, and therefore, can be substantiated experimentally.

\end{abstract}


\maketitle

\section{Introduction}

The promise of new technological breakthroughs has been a major driving
force for studying transport in meso- and nano-structures whose dimensions
are comparable to and even smaller than the mean free paths or wavelengths
of electrons~\cite{datta1,imry1,datta2}. Simultaneously, inspection  of 
electronic transport in low-dimensional systems comprising simple and 
complex structures has brought up several new underlying questions. The 
progress in experimental techniques has allowed for systematic 
investigations of artificially made nanostructures whose transport 
properties are affected or even governed by quantum effects and this makes 
it possible to perform experiments that directly probe quantum properties 
of phase coherent many-body systems.

The appearance of circular currents, induced by external magnetic fields 
in isolated (no source and drain electrodes) quantum rings, commonly known 
as persistent currents, is an astonishing quantum effect which reveals the 
significance of phase coherence of electronic wave functions in 
low-dimensional quantum systems. The phenomenon of persistent current in
normal metal rings in presence of Aharonov-Bohm (AB) flux $\phi$ has been
first exposed~\cite{yang} in the early $60$'s, and then, in $1983$ 
B\"{u}ttiker {\em et al.}~\cite{butt1} have successfully revived it and 
they have established that an isolated normal metal mesoscopic ring 
threaded by an AB flux $\phi$ carries an equilibrium current which does 
not decay over time and circulates within the sample. Following this 
pioneering work, interest in this subject has rapidly picked up with 
substantial theoretical~\cite{gefen,ambe,schm1,schm2,bary,maiti4,maiti5,
maiti6,maiti7,maiti9,skm1,skm2,skm3,skm4} and 
experimental~\cite{levy,jari,bir,chand} works. Still, many open questions 
persist in this particular issue. For instance, persistent current examined 
in disordered rings is considerably larger than the corresponding 
theoretical predictions~\cite{chand,mailly1,mailly}. In $2009$, Bluhm 
{\em et al.}~\cite{blu} have made {\em in situ} measurements using 
scanning SQUID microscope for studying magnetic properties of $33$ discrete 
mesoscopic gold rings, taking one ring at a time. Their experimental 
results fit reasonably well with the theoretically predicted 
value~\cite{gefen} only in an ensemble of $16$ nearly ballistic 
rings~\cite{mailly1} and in a single ballistic ring~\cite{mailly}. But, the
current amplitudes in single isolated diffusive gold rings~\cite{chand} 
are still order of magnitude larger than the theoretical predictions. In 
presence of electron-electron (e-e) interaction and disorder an explanation 
has been proposed~\cite{ambe,smt,mont,ph}, based on a perturbative 
calculation, which reveals persistent currents with greater amplitude 
compared to the non-interacting case, but still off by an order of 
magnitude. Moreover, the origin of the e-e interaction parameters taken 
into account in the theory is not suitably unraveled. Thus, it demands 
further studies to resolve these controversial issues.

Another challenging topic is the possible existence of a spin 
current~\cite{qf1} in mesoscopic rings with spin-orbit (SO) interaction, 
even in the absence of a magnetic field. This phenomenon may be observed 
through the recently developed Doppler and spin modulation relaxation 
techniques~\cite{vla,and1,and2}. The SO interaction is a rudimentary 
mechanism that is manifested in several fascinating properties that 
pertain to the anticipations of semiconducting structures as potential 
quantum devices. In conventional semiconducting materials two typical
SO interactions are encountered. One is called as Rashba spin-orbit 
interaction (RSOI)~\cite{rash} and the other is named as Dresselhaus 
spin-orbit interaction (DSOI)~\cite{dress}. The previous one is associated 
with electric field that is generated from the structural inversion 
asymmetry at interfaces, while the later results from the bulk inversion 
asymmetry~\cite{gini}. An additional contribution can also arise from 
surface anisotropies~\cite{prem} together with simple Rashba SO interaction, 
which is associated with the interfacial electric field normal to the 
surface that results from the band offset at the interface of two different 
semiconductors. Quantum rings - ring-shaped quantum wells fabricated at 
such heterojunctions - comprise such anisotropies are exemplary candidates 
for examining SO coupling effects in persistent currents. Note that if one 
of the components that make the interface is characterized by bulk 
asymmetry, a corresponding contribution of the DSOI will also exist at 
the interface~\cite{wnk}. In such quantum rings electronic transport will 
exhibit the interplay between these different contributions to the SO 
coupling at the interface. A sizable amount of related theoretical work 
has already revealed the distinctive features of persistent charge and 
spin currents in mesoscopic rings subjected to Rashba and Dresselhaus SO 
fields~\cite{qf1,sng,spl,loss}, however a well defined methodology for 
the prediction of persistent current in large samples is still missing 
and the magneto-transport properties of such structures are fascinating 
and remain controversial. The accurate determination of persistent current 
in such systems, in the presence of an AB flux, is a route for analyzing 
its magnetic properties.

A clear understanding of the role played by SO interactions in the phenomena 
of persistent charge and spin currents necessiates proper estimation of the
strength of these interactions. The Rashba SO interaction which is 
controlled by an external gate voltage placed in the vicinity of the 
sample~\cite{gini,prem} can be determined by the structure of the interface.
This yields, in principle, a wide range of possible values of RSOI and 
its determination in any given material is crucial~\cite{nit}. The 
feasible routes of measuring the strength of DSOI are mainly based on the 
photo-galvanic methods~\cite{yi}, measurement of electrical conductance of 
nano-wires~\cite{sc}, and an optical monitoring of the spin precession of 
the electrons~\cite{stu}. A unitary transformation has been 
explored~\cite{eg,zh} which brings out a hidden symmetry, when applied to 
the SO Hamiltonian, that has been used to establish that by making the 
strengths of the two SO interactions equal one achieves a zero spin current 
in the material~\cite{maiti1}, and this vanishing spin current is a robust 
effect which is observed even in presence of disorder~\cite{maiti1}, and 
thus, can be established experimentally. Observing the persistent charge 
current~\cite{maiti2} one can estimate the strength of DSOI, and, by 
monitoring the vanishing of persistent spin current one can determine 
both the RSOI and DSOI in a single mesoscopic ring~\cite{maiti1,maiti3}. 

The established approach to the determination of persistent 
charge~\cite{gefen,spl,maiti2,bouz,giam,yu,maiti8,maiti10,skm5} and 
spin~\cite{qf1,spl,qf2,maiti3} currents in isolated conducting rings 
is based on the evaluation of eigenvalues and eigenvectors of the system 
Hamiltonian. For large size rings such an approach becomes highly 
numerically unreliable, and most importantly - hard to speculate in 
presence of interaction with external baths. Here we propose a new 
approach, based on Green's function formalism, that circumvents the 
need to evaluate system eigenvalues and eigenfunctions. In particular, 
this Green's function methodology for determining persistent currents 
should give us access to evaluation of the magnetic properties of large 
conducting rings as well as molecular rings encountered in biopolymers. 
We firmly believe that the Green's function technique will yield 
persistent charge and spin currents a very high degree of accuracy, and 
this will definitely make it possible to consider the interplay between 
molecular structure and geometry and the resulting persistent currents 
obtained in the presence of an AB flux $\phi$ and SO interactions.

The rest of the paper is organized as follows. In Section II, the model 
quantum system and the calculation method are described. In Section III,
the numerical results are presented which describe the (i) behavior of
persistent charge current, (ii) characteristic features of persistent spin
current, and (iii) possible route of estimating the strength of RSOI and
DSOI in a single mesoscopic ring. Finally, in Section IV, we summarize our
essential results.

\section{Theoretical framework}

\subsection{Model and Hamiltonian}

\begin{figure}[ht]
{\centering \resizebox*{7.2cm}{3cm}{\includegraphics{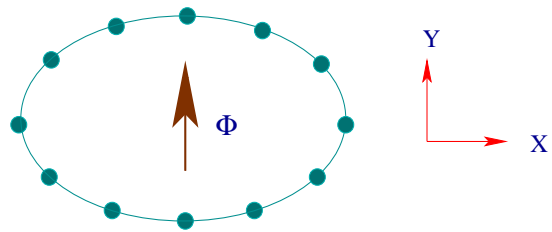}}\par}
\caption{(Color online). Schematic representation of the model quantum
system where a mesoscopic ring is threaded by an AB flux $\phi$. The filled
circles correspond to the positions of the atomic sites.}
\label{ring}
\end{figure}

We consider a mesoscopic ring which is subjected to both Rashba and 
Dresselhaus SO fields. The ring is threaded by an AB flux $\phi$ which
is measured in unit of $\phi_0=ch/e$, the elementary flux-quantum. A
schematic view of this ring is illustrated in Fig.~\ref{ring}. 
The TB Hamiltonian of such a $N$-site ring in the site representation 
reads as~\cite{maiti1,maiti2,maiti3},
\begin{eqnarray}
\mbox{\boldmath $H$}_{\mbox{\tiny R}}&=&\sum_n \mbox{\boldmath$c_n^{\dagger} 
\epsilon_n c_n$} + \sum_n \left(\mbox{\boldmath $c_{n+1}^{\dagger} t$} 
\,e^{i \theta} \mbox {\boldmath $c_n$} + h.c. \right) \nonumber \\
&-& \sum_n \left(\mbox{\boldmath $c_{n+1}^{\dagger} 
($}i \mbox{\boldmath$\sigma_x)\alpha$} \cos\varphi_{n,n+1} \,e^{i \theta} 
\mbox{\boldmath $c_n$} + h.c. \right) \nonumber \\
& - & \sum_n \left(\mbox{\boldmath $c_{n+1}^{\dagger} 
($}i \mbox{\boldmath$\sigma_y)\alpha$} \sin\varphi_{n,n+1} \,e^{i \theta} 
\mbox{\boldmath $c_n$} + h.c. \right) \nonumber \\
& + & \sum_n \left(\mbox{\boldmath $c_{n+1}^{\dagger} 
($}i \mbox{\boldmath$\sigma_y)\beta$} \cos\varphi_{n,n+1} \,e^{i \theta} 
\mbox{\boldmath $c_n$} + h.c. \right) \nonumber \\
& + & \sum_n \left(\mbox{\boldmath $c_{n+1}^{\dagger} 
($}i \mbox{\boldmath$\sigma_x)\beta$} \sin\varphi_{n,n+1} \,e^{i \theta} 
\mbox{\boldmath $c_n$} + h.c. \right),
\label{equ1}
\end{eqnarray}
where, $\theta=2\pi \phi/N$ is the phase factor associated with the
hopping of an electron between nearest-neighbor sites in presence of 
the AB flux $\phi$ and $\varphi_{n,n+1}=(\varphi_n+\varphi_{n+1})/2$, 
where $\varphi_n=2\pi(n-1)/N$. The other factors are defined as
follows. \\
\mbox{\boldmath $c_n^{\dagger}$}=$\left(\begin{array}{cc}
c_{n \uparrow}^{\dagger} & c_{n \downarrow}^{\dagger} 
\end{array}\right),$
\mbox{\boldmath $c_n$}=$\left(\begin{array}{c}
c_{n \uparrow} \\
c_{n \downarrow}\end{array}\right),$
\mbox{\boldmath $\epsilon_n$}=$\left(\begin{array}{cc}
\epsilon_{n\uparrow} & 0 \\
0 & \epsilon_{n\downarrow} \end{array}\right),$ \\ 
\mbox{\boldmath $t$}=$t\left(\begin{array}{cc}
1 & 0 \\
0 & 1 \end{array}\right),$
\mbox{\boldmath $\alpha$}=$\left(\begin{array}{cc}
\alpha & 0 \\
0 & \alpha \end{array}\right),$
\mbox{\boldmath $\beta$}=$\left(\begin{array}{cc}
\beta & 0 \\
0 & \beta \end{array}\right),$ \\ 
\mbox{\boldmath $\sigma_{x}$}=$\left(\begin{array}{cc}
0 & 1 \\
1 & 0 \end{array}\right),$
\mbox{\boldmath $\sigma_{y}$}=$\left(\begin{array}{cc}
0 & -i \\
i & 0 \end{array}\right),$
\mbox{\boldmath $\sigma_{z}$}=$\left(\begin{array}{cc}
1 & 0 \\
0 & -1 \end{array}\right)$, \\
~\\
\noindent
where, $c_{n \sigma}^{\dagger}$ and $c_{n \sigma}$ are the creation and 
annihilation operators, respectively, for an electron with spin $\sigma$ 
($\uparrow,\downarrow$) at $n$-th site. The nearest-neighbor hopping
integral is described by the parameter $t$ and $\epsilon_{n\sigma}$ denotes 
the on-site energy of an electron at the site $n$ of the ring with spin 
$\sigma$. The factors $\alpha$ and $\beta$ corresponds to the strengths 
of Rashba and Dresselhaus SO fields, respectively, and 
\mbox{\boldmath $\sigma_x$}, \mbox{\boldmath $\sigma_y$} and 
\mbox{\boldmath $\sigma_z$} are the conventional Pauli spin matrices. 

\subsection{Persistent charge current}

At absolute zero temperature ($T=0\,$K), net persistent charge current
carried by a ring for a particular electron filling is obtained from the
expression,
\begin{equation}
I_c(\phi)=\int\limits_{-\infty}^{\mu} J_c(E)\, dE,
\label{equ2}
\end{equation}
where, $\mu$ describes the chemical potential of the ring and $J_c(E)$ 
represents the charge current density. In terms  of Green's functions 
(see Appendix~\ref{pc} and Appendix~\ref{pcso} for comprehensive 
derivations) it ($J_c(E)$) gets the form:
\begin{eqnarray}
J_c(E) &=& -\frac{e}{N}\sum_n 
\left\{t\left(\mbox{\boldmath$G$}_{n+1\uparrow,n\uparrow}^r
-\mbox{\boldmath$G$}_{n+1\uparrow,n\uparrow}^a\right)e^{-i\theta} \right.
\nonumber \\
 & & \left. - t\left(\mbox{\boldmath$G$}_{n\uparrow,n+1\uparrow}^r
-\mbox{\boldmath$G$}_{n\uparrow,n+1\uparrow}^a\right)e^{i\theta}\right\}
\nonumber \\
 & & -\frac{e}{N}\sum_n 
\left\{t\left(\mbox{\boldmath$G$}_{n+1\downarrow,n\downarrow}^r
-\mbox{\boldmath$G$}_{n+1\downarrow,n\downarrow}^a\right)e^{-i\theta} \right.
\nonumber \\
 & & \left. - t\left(\mbox{\boldmath$G$}_{n\downarrow,n+1\downarrow}^r
-\mbox{\boldmath$G$}_{n\downarrow,n+1\downarrow}^a\right)e^{i\theta}\right\} 
\nonumber \\
 & & -\frac{e}{N}\sum_n \left(i \alpha e^{-i\varphi_{n,n+1}}-\beta
e^{i\varphi_{n,n+1}} \right) \nonumber \\
 & & \times \left(\mbox{\boldmath$G$}_{n+1\downarrow,n\uparrow}^r
-\mbox{\boldmath$G$}_{n+1\downarrow,n\uparrow}^a\right)e^{-i\theta} 
\nonumber \\
 & & -\frac{e}{N}\sum_n \left(i \alpha e^{i\varphi_{n,n+1}}+\beta
e^{-i\varphi_{n,n+1}} \right) \nonumber \\
 & & \times \left(\mbox{\boldmath$G$}_{n\uparrow,n+1\downarrow}^r
-\mbox{\boldmath$G$}_{n\uparrow,n+1\downarrow}^a\right)e^{i\theta} 
\nonumber \\
 & & -\frac{e}{N}\sum_n \left(i \alpha e^{i\varphi_{n,n+1}}+\beta
e^{-i\varphi_{n,n+1}} \right) \nonumber \\
 & & \times \left(\mbox{\boldmath$G$}_{n+1\uparrow,n\downarrow}^r
-\mbox{\boldmath$G$}_{n+1\uparrow,n\downarrow}^a\right)e^{-i\theta} 
\nonumber \\
 & & -\frac{e}{N}\sum_n \left(i \alpha e^{-i\varphi_{n,n+1}}-\beta
e^{i\varphi_{n,n+1}} \right) \nonumber \\
 & & \times \left(\mbox{\boldmath$G$}_{n\downarrow,n+1\uparrow}^r
-\mbox{\boldmath$G$}_{n\downarrow,n+1\uparrow}^a\right)e^{i\theta}, 
\label{chcurr}
\end{eqnarray}
where, $\mbox{\boldmath$G$}^r$ is the retarded Green's function 
defined as $\mbox{\boldmath$G$}^r=\left(\mbox{\boldmath$E$}-
\mbox{\boldmath $H$}_{\mbox{\tiny R}} + 
i \eta \mbox{\boldmath$I$}\right)^{-1}$ with $\eta \rightarrow 0^+$, 
and, $\mbox{\boldmath$G$}^a=\left(\mbox{\boldmath$G$}^r\right)^{\dagger}$.  
$\mbox{\boldmath$I$}$ being the identity matrix.

\subsection{Persistent spin current}

Similar to Eq.~\ref{equ2}, we determine polarized spin current along the
quantized direction ($+Z$) at absolute zero temperature ($T=0\,$K) from 
the relation,
\begin{equation}
I_s(\phi)=\int\limits_{-\infty}^{\mu} J_s(E)\, dE,
\label{equ4}
\end{equation}
where, $J_s(E)$ corresponds to the spin current density and it becomes
(see Appendix~\ref{psso} for complete derivations),
\begin{eqnarray}
J_s(E) &=& -\frac{1}{N}\sum_n \left\{t\left(\mbox{\boldmath$G$}_{n+1\uparrow,
n\uparrow}^r-\mbox{\boldmath$G$}_{n+1\uparrow,n\uparrow}^a\right)e^{-i\theta}
\right. \nonumber \\ 
& & \left. - t\left(\mbox{\boldmath$G$}_{n\uparrow,n+1\uparrow}^r
-\mbox{\boldmath$G$}_{n\uparrow,n+1\uparrow}^a\right)e^{i\theta}\right\}
\nonumber \\
 & & +\frac{1}{N}\sum_n \left\{t\left(\mbox{\boldmath$G$}_{n+1\downarrow,
n\downarrow}^r-\mbox{\boldmath$G$}_{n+1\downarrow,n\downarrow}^a\right)
e^{-i\theta} \right. \nonumber \\
& & \left. - t\left(\mbox{\boldmath$G$}_{n\downarrow,n+1\downarrow}^r
-\mbox{\boldmath$G$}_{n\downarrow,n+1\downarrow}^a\right)e^{i\theta}\right\}.
\label{spcurr}
\end{eqnarray}
Thus, introducing the notion of persistent charge and spin
current densities $J_c(E)$ and $J_s(E)$ in terms of the retared and 
advanced Green's functions, expressed in Eqs.~\ref{chcurr} and \ref{spcurr} 
respectively, we eventually determine persistent charge and spin currents 
by integrating the current densities over a suitable energy window (see 
Eqs.~\ref{equ2} and \ref{equ4}) associated with the electron filling. The 
detailed and long calculations of these charge and spin current 
densities as a function of retared and advanced Green's functions are 
presented in the Appendices~\ref{pc}-\ref{psso}, to have a complete idea
for calculating the desired quantities. Our new approach, the so-called 
Green's function approach, clearly suggests how to circumvent the need
to evaluate system eigenvalues and eigenfunctions for evaluating persistent
currents, as used in conventional methods. Here, it should be noted that 
the present scheme is well applicable both for the ordered and disordered
systems since all the mathematical expressions are exactly invariant in both
these two cases. Only the magnitudes of different elements of the Green's 
functions get changed depending on impurity strength of the sample. This
behavior essentially demands the robustness of the present new technique.

\section{Numerical results and discussion}

In what follows, we will present numerical results computed for circulating
charge and spin currents in mesoscopic rings based on Green's function
formalism. In all calculations we measure the energy scale in unit of the
hopping integral $t$ which is set equal to $1$. The Rashba and Dresselhaus 
SO coupling strengths are also scaled in unit of this hopping parameter $t$. 
Throughout the numerical analysis we restrict ourselves to absolute zero 
temperature and fix $c=h=e=1$.

First, we focus on the impurity-free mesoscopic rings, and, for such a ring 
we put $\epsilon_{n \uparrow}=\epsilon_{n \downarrow}=0$ for all $n$ in the
TB Hamiltonian Eq.~\ref{equ1}. 

\begin{figure}[ht]
{\centering \resizebox*{8cm}{13cm}
{\includegraphics{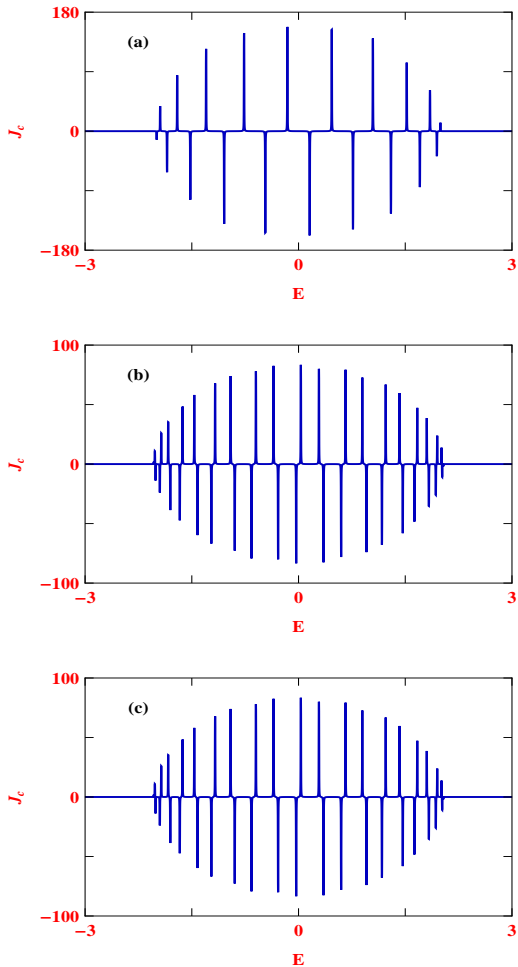}}\par}
\caption{(Color online). Persistent charge current density as a function
of energy for a $20$-site ordered ring setting $\phi=\phi_0/4$, where
(a) $\alpha=\beta=0$, (b) $\alpha=0.2$ and $\beta=0$, and (c) $\alpha=0$
and $\beta=0.2$.}
\label{chargecurrden}
\end{figure}

In Fig.~\ref{chargecurrden} we present the variation of persistent
charge current density $J_c$ as a function of energy $E$ for some typical
values of Rashba and Dresselhaus spin-orbit fields. The results are
computed for a $20$-site impurity-free mesoscopic ring when the AB flux
$\phi$ is set at $\phi_0/4$. For a better viewing of distinct peaks in
the density spectrum here we display the results for such a small size
ring. Several interesting patterns are obtained those can be analyzed as
follows. From the spectra it is observed that the charge current density
exhibits sharp peaks and dips for some particular energy values, while
it drops to zero for other energies. All these peaks and dips are associated
with the energy eigenvalues of the ring.

For the particular case when the ring is free from any kind of SO
interaction and subjected to a non-zero magnetic flux, apart from integer 
or half-integer multiples of the elementary flux-quantum, the energy levels 
are two-fold degenerate which results in total $20$ peaks and dips in the 
current density spectrum (see Fig.~\ref{chargecurrden}(a)). It is quite
interesting to note that the peaks and dips appear alternately throughout
the band spectrum  which essentially leads to an important conclusion that 
successive energy levels carry currents in opposite directions. This behavior
suggests the vanishing net current for the complete band filling. 
\begin{figure}[ht]
{\centering \resizebox*{8cm}{13cm}
{\includegraphics{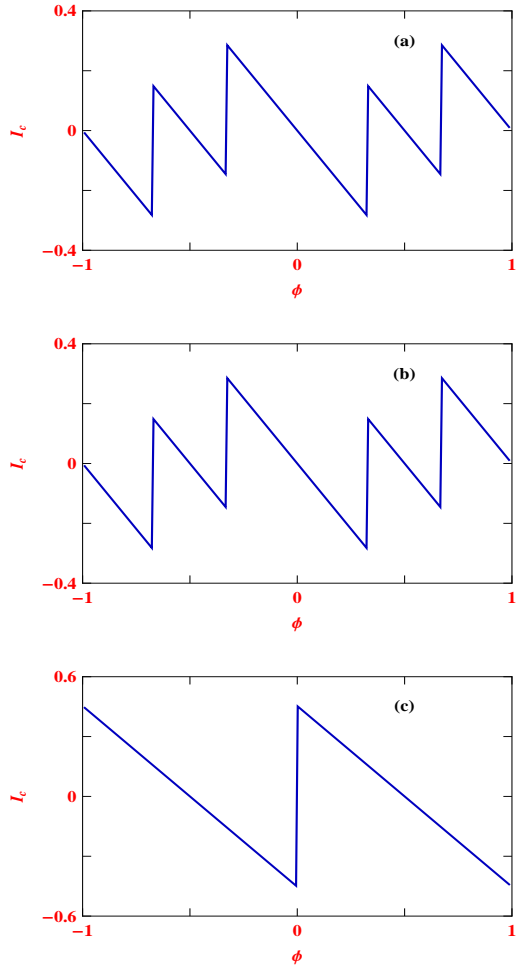}}\par}
\caption{(Color online). Persistent charge current as a function of
flux $\phi$ for a $60$-site ordered ring, where (a) $\alpha=0.3$ and
$\beta=0$, (b) $\alpha=0$ and $\beta=0.3$, and (c) $\alpha=\beta=0.3$.
The chemical potential $\mu$ is fixed at $0$.}
\label{chargecurrent}
\end{figure}
Furthermore, one can also utilize this charge current density spectrum to
predict the nature of extendedness of different energy levels by 
superimposing the average density of states (ADOS) on it. A non-zero
contribution, viz, a peak or a dip, to the charge current will be obtained
from the conducting states, while it becomes zero for the localized ones.
This is another way of estimating the localization phenomenon in addition
to the conventional methodologies~\cite{maiti11,maiti12,maiti13}.

The charge current density spectrum gets significantly modified when
we include SO coupling in this AB ring. The results are shown in
Figs.~\ref{chargecurrden}(b) and (c), where in (b) we set a finite value
of RSOI strength keeping the strength of DSOI as zero, while in (c) these
parameter values get interchanged. It shows that the total number of 
peaks in the current density profiles, Figs.~\ref{chargecurrden}(b) and (c),
associated with energy levels of the ring becomes exactly twice compared
to the interacting free mesoscopic ring, viz., $\alpha=\beta=0$ (see 
Fig.~\ref{chargecurrden}(a)). This is because of the complete removal of 
degenerate energy eigenstates of the AB ring subjected to SO interaction. 
One more important property is also observed from these spectra 
(Figs.~\ref{chargecurrden}(b) and (c)) that the magnitude and sign of 
persistent charge current density $J_c$ for any energy window when 
the AB ring is subjected to RSOI only (Fig.~\ref{chargecurrden}(b)) are
exactly identical to the ring described with only DSOI
(Fig.~\ref{chargecurrden}(c)). Thus, it should be emphasized that the phase 
reversal in charge current density does not take place by interchanging 
the role played by $\alpha$ and $\beta$ into the Hamiltonian Eq.~\ref{equ1}. 
This phenomenon can be implemented from the following analytical prescription.

The Rashba and Dresselhaus SO interaction terms, called as, 
\mbox{\boldmath $H_{\mbox{\tiny RSOI}}$} 
and \mbox{\boldmath $H_{\mbox{\tiny DSOI}}$} in the TB Hamiltonian 
Eq.~\ref{equ1} can be transformed into each other through a simple 
relation: 
$\mbox{\boldmath $U^{\dag} H_{\mbox{\tiny RSOI}} U$}$ = 
$\mbox{\boldmath $H_{\mbox{\tiny DSOI}}$}$. Here, 
$\mbox{\boldmath $U$}$ = $\left(\mbox{\boldmath$\sigma$}_x + 
\mbox{\boldmath$\sigma$}_y \right)/\sqrt{2}$ is the unitary matrix. Thus,
any energy eigenstate $|\mathcal{M}^{\prime}\rangle$ of the transformed
Hamiltonian \mbox{\boldmath $H_{\mbox{\tiny DSOI}}$} can be demonstrated 
in terms of the eigenstate $|\mathcal{M}\rangle$ of the Hamiltonian
\mbox{\boldmath $H_{\mbox{\tiny RSOI}}$} through the relation
$|\mathcal{M}^{\prime} \rangle$=$\mbox{\boldmath $U$} |\mathcal{M}\rangle$.
This transformation leads to the charge current for the ring with only
Dresselhaus SO coupling as,
\begin{eqnarray}
J_c^m|_{\mbox{\tiny DSOI}} & = & 
\langle\mathcal{M}^{\prime}|\mbox{\boldmath $J$}_c|\mathcal{M}^{\prime}\rangle 
\nonumber \\
& = & \langle\mathcal{M}|\mbox{\boldmath $U$}^\dag \mbox{\boldmath$J$}_c 
\mbox{\boldmath $U$}|\mathcal{M}\rangle \nonumber \\
& = & \langle\mathcal{M}|\mbox{\boldmath $U$}^\dag \frac{e}{N}
\mbox{\boldmath$\dot{x}$}\mbox{\boldmath $U$}|\mathcal{M}\rangle \nonumber \\
& = & \langle\mathcal{M}| \frac{e}{N}\mbox{\boldmath$\dot{x}$}|\mathcal{M}
\rangle \nonumber \\
& = & J_c^m|_{\mbox{\tiny RSOI}}.
\label{eqs1}
\end{eqnarray}
The above expression clearly establishes the reason for not affecting the 
sign and magnitude of charge current density upon the exchange of the 
role played by RSOI and DSOI.

Once the charge current density, $J_c(E)$, is evaluated using the relation 
presented in Eq.~\ref{chcurr}, the net persistent charge current $I_c$ in 
the ring can be easily determined by integrating the density spectrum 
(see Eq.~\ref{equ2}) up to a certain energy range depending on the electron 
filling. As illustrative example, in Fig.~\ref{chargecurrent} we present 
the variation of persistent charge current as a function of magnetic flux
$\phi$ for an ordered ring considering different values of the SO coupling 
strengths. Here we set $N=60$ and $\mu=0$. The effect of the Rashba SO
coupling is examined in the spectrum Fig.~\ref{chargecurrent}(a), setting
the Dresselhaus SO interaction to zero. It is observed that the persistent
charge current exhibits kink-like structures together with phase reversals
at several values of flux $\phi$, which are however in general not unusual
even in the absence of any SO coupling. These are essentially due to the 
band crossing in energy spectra and are immensely sensitive to the filling 
factor $\mu$. Though we have computed charge currents for different band 
fillings through ample numerical calculations, here we present results 
for a particular electron filling, as a test example, to establish our 
Green's function approach for the estimation of persistent current in a 
\begin{figure}[ht]
{\centering \resizebox*{8cm}{13cm}
{\includegraphics{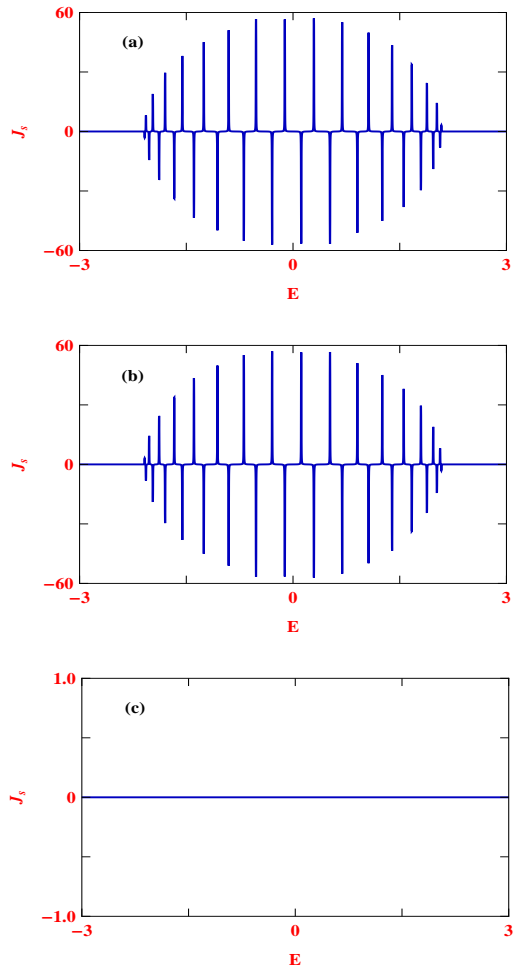}}\par}
\caption{(Color online). Persistent spin current density as a function of
energy for a $16$-site ordered ring, where (a) $\alpha=0.3$ and $\beta=0$,
(b) $\alpha=0$ and $\beta=0.3$, and (c) $\alpha=\beta=0.3$. Here, we
set $\phi=\phi_0/4$.}
\label{spincurrden}
\end{figure}
mesoscopic ring. In addition to the above issues it is also very important 
to point out that, for a particular value of the AB flux $\phi$ the current 
amplitude strongly depends on the chemical potential of the sample. An 
exhaustive analysis has already been given by Splettstoesser 
{\em et al.}~\cite{spl} in this line. 

Figure~\ref{chargecurrent}(b) illustrates the situation in which the
ring is described with DSOI only i.e., the other SO coupling term (RSOI) 
is set equal to zero. Since, the reversal of the roles governed by the 
variables $\alpha$ and $\beta$ into the TB Hamiltonian Eq.~\ref{equ1} does
not anyway alter the physical picture of charge current density spectrum,
the current-flux characteristics for the rings described with RSOI only 
(see Fig.~\ref{chargecurrent}(a)) become exactly identical to those with 
the rings in presence of DSOI only (see Fig.~\ref{chargecurrent}(b)). This 
phenomenon can also be justified from our analytical arguments presented
in Eq.~\ref{eqs1}.

The combined outcome of both these two SO fields on persistent charge 
current is scrutinized in Fig.~\ref{chargecurrent}(c), where we specify 
$\alpha=\beta=0.3$. The other parameters are kept unchanged as taken in 
Figs.~\ref{chargecurrent}(a) and (b). In presence of both these two
\begin{figure}[ht]
{\centering \resizebox*{8cm}{13cm}
{\includegraphics{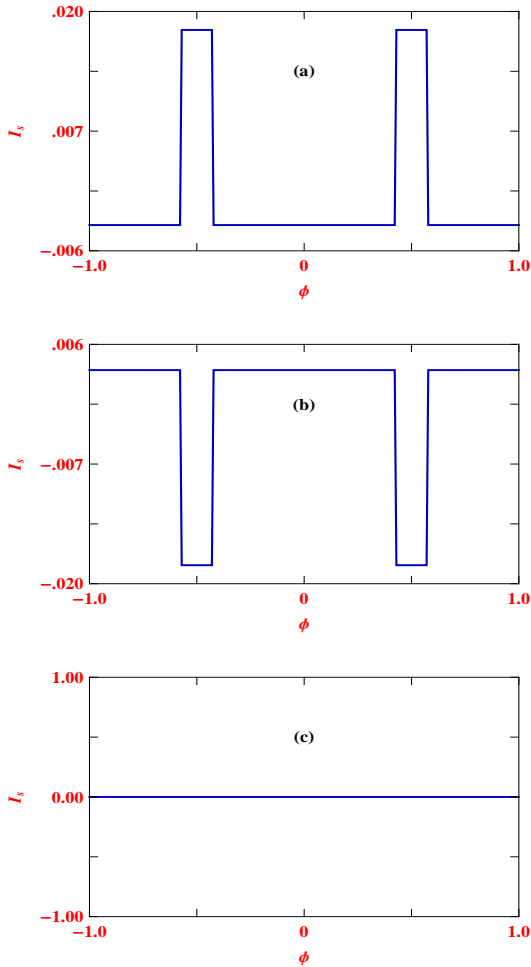}}\par}
\caption{(Color online). Persistent spin current as a function of flux
$\phi$ for a $100$-site ordered ring when the chemical potential $\mu$ is
fixed at $0$, where (a) $\alpha=0.4$ and $\beta=0$, (b) $\alpha=0$ and
$\beta=0.4$, and (c) $\alpha=\beta=0.4$.}
\label{spincurrent}
\end{figure}
interactions, the nature of persistent current for different values of 
$\phi$ changes appreciably compared to the case when only one SO 
interaction is present. This is due to the fact that the inclusion of 
both these two SO fields reforms the electronic band structure of the 
ring and thus strongly affects the pattern of the circulating current. In 
short, it can be emphasized that, charge current is distinctly sensitive 
to the SO coupling strength and magnetic flux threaded by the ring. All 
these currents vary repeatedly bearing $\phi_0$ ($=1$, in our choice 
of units where $c=e=h=1$) flux-quantum periodicity.

Now, we extend our discussion on persistent spin current in SO interaction
induced impurity-free mesoscopic rings, where currents are computed from
the Green's function formalism. Before addressing these results, we first 
analyze the behavior of persistent spin current density, determined from 
the relation given in Eq.~\ref{spcurr}, to make this communication a self 
contained study. In Fig.~\ref{spincurrden} the energy dependent spin 
current density spectra are presented for a $16$-site ordered ring 
considering $\phi=\phi_0/4$, where the upper, middle and lower panels 
correspond to the results for three different set of parameter values of 
$\alpha$ and $\beta$. Interestingly, we see that the spin current density
in the ring described with only Dresselhaus SO coupling (see 
Fig.~\ref{spincurrden}(b)) changes its sign keeping the magnitude unaltered 
compared to the ring with RSOI only (see Fig.~\ref{spincurrden}(a)). This
sign reversal behavior can be viewed as follows. Using the similar 
prescription presented in Eq.~\ref{eqs1}, the spin current for the ring
described with only DSOI gets the form,
\begin{eqnarray}
J_s^m|_{\mbox{\tiny DSOI}} & = & 
\langle\mathcal{M}^{\prime}|\mbox{\boldmath $J$}_s|\mathcal{M}^{\prime}\rangle 
\nonumber \\
& = & \langle\mathcal{M}|\mbox{\boldmath $U$}^\dag \mbox{\boldmath$J$}_s 
\mbox{\boldmath $U$}|\mathcal{M}\rangle \nonumber \\
& = & \langle\mathcal{M}|\mbox{\boldmath $U$}^\dag \frac{1}{2N}
\left( \mbox{\boldmath$\sigma_z \dot{x}$} + 
\mbox{\boldmath$\dot{x} \sigma_z$} \right)
\mbox{\boldmath $U$}|\mathcal{M}\rangle \nonumber \\
& = & \langle\mathcal{M}| \frac{1}{2N} 
\left(-\mbox{\boldmath$\sigma_z \dot{x}$}-\mbox{\boldmath$\dot{x} 
\sigma_z$} \right)|\mathcal{M}
\rangle \nonumber \\
& = & -J_s^m|_{\mbox{\tiny RSOI}}.
\label{eqs2}
\end{eqnarray}
This relation clearly describes the sign reversal of spin current density 
upon the interchange of the parameters $\alpha$ and $\beta$ into the 
Hamiltonian (Eq.~\ref{equ1}) of the ring. From this analysis we can also
justify the vanishing nature of persistent spin current density for the
entire density spectrum (see Fig.~\ref{spincurrden}(c)) when the Dresselhaus 
SO interaction strength becomes precisely identical to the strength of the 
Rashba term. This is an interesting observation and may lead to a possible
route for estimating the strength of anyone of the SO fields provided the
other one is known. A detailed analysis of it can be obtained from the
following current-flux characteristics.

Similar to persistent charge current, we also compute net persistent spin
current $I_s$ in the ring by integrating the spin current density $J_s(E)$
(see Eq.~\ref{equ4}) over a finite energy range associated with the filling 
factor $\mu$. As representative example, in Fig.~\ref{spincurrent} we show
the variation of persistent spin current as a function of AB flux $\phi$
for a $100$-site ordered ring when the chemical potential $\mu$ is set at
zero. Figure~\ref{spincurrent}(a) illustrates the situation in which the
ring is described with RSOI only, while in Figure~\ref{spincurrent}(b) the
effect of Dresselhaus SO coupling on spin current is presented. From these
spectra (Figs.~\ref{spincurrent}(a) and (b)) we observe that, spin current 
in the ring characterized with DSOI only alters its sign keeping the 
magnitude unchanged compared to the ring with Rashba term only, which is
however exactly what we expect from the spin current density spectra shown
in Fig.~\ref{spincurrden}, since current is computed by integrating the
density function $J_s$. The usual phase reversals at several values of AB 
flux $\phi$ associated with the band crossing in energy spectra together 
with $\phi_0$ flux-quantum periodicity are also noticed from these 
current-flux characteristics.

Certainly, whenever the strength of Dresselhaus SO coupling becomes exactly 
identical to that of Rashba SO interaction, spin current becomes zero for 
the entire flux window. It is shown in Fig.~\ref{spincurrent}(c), where we 
set $\alpha=\beta=0.4$. This vanishing nature of persistent spin current is 
\begin{figure}[ht]
{\centering \resizebox*{8cm}{13cm}
{\includegraphics{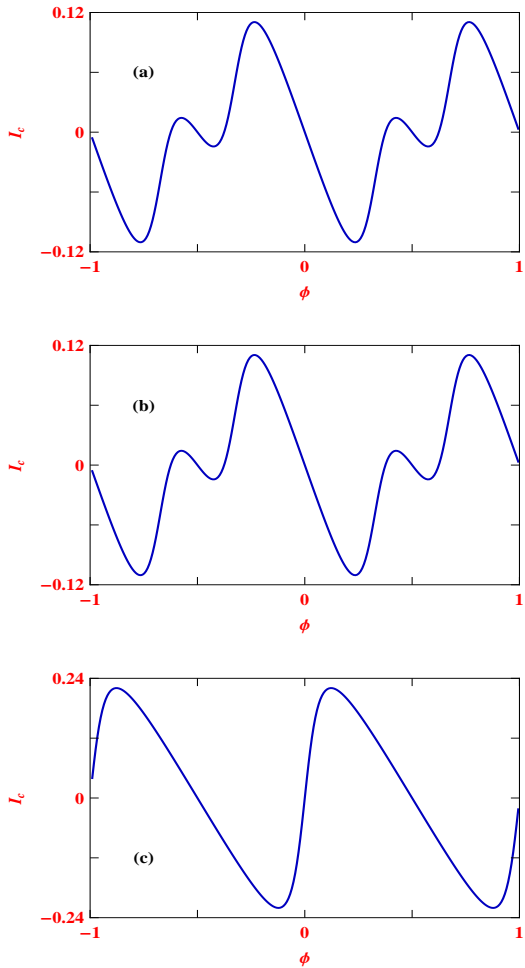}}\par}
\caption{(Color online). Persistent charge current as a function of flux
$\phi$ for a $60$-site ring in presence of disorder ($W=1$) for the same
parameter values used in Fig.~\ref{chargecurrent}.}
\label{chargedis}
\end{figure}
detected for any non-zero value of Rashba SO interaction provided it becomes 
equal to the Dresselhaus term, and also this behavior is independent of the 
band filling which we establish
through our vast numerical calculations. This phenomenon, in principle,
gives a possibility of estimating anyone of the SO fields if the other one
is known. By means of an outside gate voltage one can control the Rashba
SO coupling, and thus its strength can be determined. This suggests that, 
monitoring the RSOI in a mesoscopic ring we will get an absolute vanishing 
spin current when the strength of the Dresselhaus SO coupling becomes 
identical to that of the RSOI. Thus, from a realizable experimental 
measurement of persistent spin current one can evaluate the strength of 
Dresselhaus SO coupling. Additionally, it is also important to state that 
one may determine the strength of Rashba term employing this same mechanism
provided the other term is known.

Up to now we have demonstrated the results for perfect rings. For more
practical implications we now focus our attention on mesoscopic rings
in presence of impurities. Impurities are introduced randomly in site
potentials ($\epsilon_{n \uparrow}$ and $\epsilon_{n \downarrow}$) i.e.,
\begin{figure}[ht]
{\centering \resizebox*{8cm}{13cm}
{\includegraphics{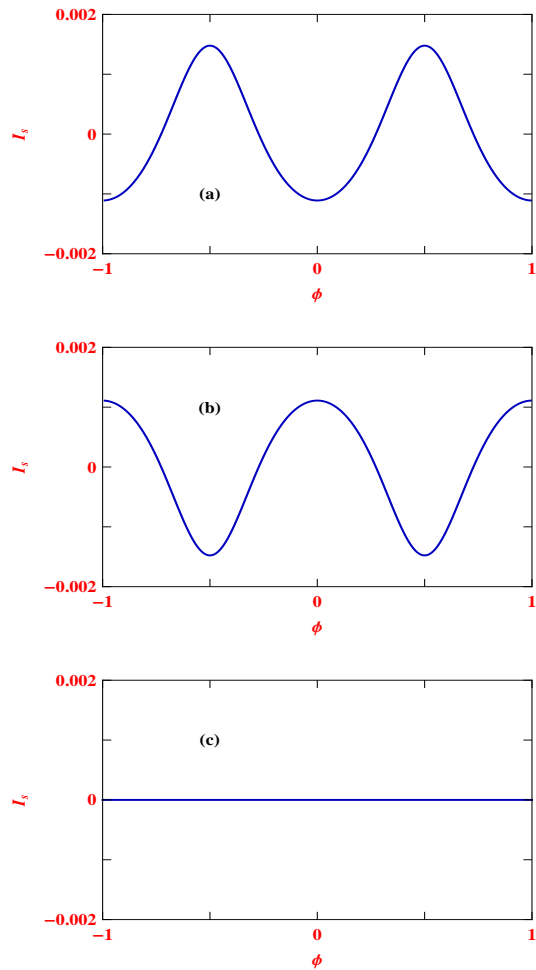}}\par}
\caption{(Color online). Persistent spin current as a function of flux
$\phi$ for a $100$-site ring in presence of disorder ($W=1$). The other
model parameters are kept unchanged as used in Fig.~\ref{spincurrent}.}
\label{spindis}
\end{figure}
diagonal disorder, through a `Box' distribution function of width $W$,
and the results averaged over $1000$ disorder configurations are presented.
In Fig.~\ref{chargedis} we present the results of persistent charge current
for a disordered ring considering $W=1$ for different values of SO coupling 
strengths. Here we set $N=60$ and the results are computed for $\mu=0$, as
a typical example. Several interesting features are obtained. Firstly, the
current shows a continuous variation with flux $\phi$. This is essentially
due to the fact that disorder makes a smooth variation of energy levels and
eliminate band crossings those are mostly observed in impurity-free
rings. The other important observation is that, in presence of impurities
the current amplitude gets suppressed compared to the ordered case which
can be clearly visible from the spectra plotted in Figs.~\ref{chargecurrent} 
and \ref{chargedis}. In presence of impurities the energy eigenstates are
localized which results a reduction of persistent current, though this 
reduction is quite small compared to the ring without any SO interaction. 
In the absence of any SO coupling, disorder suppresses current amplitude 
almost to zero which is not unfamiliar in conventional disordered rings. 
But, with the inclusion of SO interaction current increases significantly 
and becomes quite proportionate to that of a perfect ring. A detailed 
analysis behind this mechanism has already been reported in our recent 
work~\cite{maiti2}.

Finally, in Fig.~\ref{spindis} we present the results of persistent spin
current in a mesoscopic ring to give a complete exposure of our Green's 
function formalism for the evaluation of persistent current even in 
presence of impurities. Here we set $W=1$ and all the other parameters 
are kept unchanged as taken in Fig.~\ref{spincurrent}. As usual the
current varies continuously with flux $\phi$ exhibiting $\phi_0$
flux-quantum periodicity and it gets a reduced amplitude compared to the
perfectly ordered ring. All the other properties i.e., the sign reversal
upon the interchange of the role played by the parameters $\alpha$ and
$\beta$ into the Hamiltonian Eq.~\ref{equ1} and the vanishing nature of
spin current when the RSOI becomes equal to DSOI, remain exactly valid 
in the presence of impurities.

\section{Summary and outlook}

In the present work, we have proposed a new approach based on Green's
function formalism within a tight-binding framework to evaluate precisely
the persistent charge and spin currents in spin-orbit interaction induced
AB rings. The essential results are summarized as follows. 

As already pointed out that, the standard methodology to the determination of 
circulating charge and spin currents in isolated conducting loops is based
on the evaluation of eigenvalues and eigenvectors of the system Hamiltonian,
which is highly numerically unstable especially for large size rings. Not
only that it is really very hard to generalize in presence of interaction 
with external baths, if any. The present approach, the so-called Green's
function technique, circumvents the need to determine eigenvalues and 
eigenvectors of the system, and in particular, this methodology should 
give us access to predict the magnetic properties of large conducting
rings as well as molecular rings encountered in biopolymers. We strongly 
believe that the present analysis yields persistent currents a very high
degree of accuracy and leads to consider interplay of AB flux and 
geometry in magneto-transport of conducting loops subjected to Rashba and 
Dresselhaus SO fields.

It is worth pointing out that, in the present work we mainly concentrate
on the new technique for the determination of persistent charge and spin
currents. With this technique, we have also provided one possible route of
estimating the strength of anyone of the SO fields provided the other one
is known. This can be done by measuring persistent spin current which 
vanishes completely when the DSOI becomes equal to the RSOI, and this 
vanishing effect is also observed even in presence of impurities. It
essentially supports us to propose an experiment towards this direction.

Before we end, we would like to mention that though we have computed 
persistent charge and spin currents for different band fillings considering
different size rings through extensive numerical calculations, but here
we have presented our results for some typical parameter values to 
explain the physical phenomena computed from our theoretical framework.
All these physical pictures will be absolutely invariant for other parameter
values also and thus demands the robustness of this new technique. 
In a forthcoming paper, we will provide the way of determining persistent
charge and spin currents using this Green's function technique considering
the effect of electron-electron interaction.

\begin{acknowledgments}
First author is thankful to Prof. A. Nitzan for useful conversations, 
and S. Saha and P. Dutta for helpful discussions.
\end{acknowledgments}

\appendix
\section{Green's function approach for persistent charge current density
when the ring is free from SO interactions} 
\label{pc}

First, we consider the model quantum ring, presented in Fig.~\ref{ring}, 
and establish persistent charge current density in terms of Green's 
function setting $\alpha=\beta=0$. Under this condition, TB Hamiltonian 
of the ring with site energy $\epsilon_n$ and nearest-neighbor hopping
interaction $t$ becomes,
\begin{eqnarray}
\mbox{\boldmath $H$} &=& \sum_n \epsilon_n \, c_n^{\dagger} c_n + 
\sum_n \left(c_{n+1}^{\dagger} \,t \, c_n e^{i\theta} + 
c_n^{\dagger} \,t \, c_{n+1} e^{-i\theta}\right). \nonumber \\
\label{ap1}
\end{eqnarray} 
In terms of the velocity operator $\mbox{\boldmath$\dot{x}$}$ the charge 
current operator $\mbox{\boldmath$J_c$}$ can be written as,
\begin{equation}
\mbox{\boldmath $J_{c}$}=\frac{1}{N} e \mbox{\boldmath ${\dot{x}}$}=
\frac{2\pi i e}{Nh}\left[\mbox{\boldmath ${H}$},\mbox{\boldmath ${x}$}
\right],
\label{ap2}
\end{equation}
where, \mbox{\boldmath ${x}$}=$\sum\limits_n c_n^{\dagger}\, n\, c_n$ 
is the position operator. Substituting 
$\mbox{\boldmath$H$}$ and $\mbox{\boldmath$x$}$ in Eq.~\ref{ap2} and
doing a quite long but straightforward calculation we eventually reach to
the expression,
\begin{equation}
\mbox{\boldmath$J_c$}=\frac{2\pi i e}{Nh} \sum_n \left(c_n^{\dagger} \,
t\, c_{n+1} e^{-i\theta} - c_{n+1}^{\dagger} \, t \, c_n e^{i\theta}\right),
\label{ap3}
\end{equation}
and, for a particular energy eigenstate $|\mathcal{M}\rangle=\sum\limits_p 
a_p^m |p\rangle$, it leads to the persistent charge current:
\begin{eqnarray}
J_c^m &=& \langle \mathcal{M}|\mbox{\boldmath$J_c$}|\mathcal{M}\rangle 
\nonumber \\
&=&\frac{2\pi i e t}{Nh} \sum_n \left(a_n^{m\,*}a_{n+1}^m e^{-i\theta} 
-a_{n+1}^{m\,*}a_n^m e^{i\theta}\right),
\label{ap4}
\end{eqnarray}
where, $|p\rangle$'s are the Wannier states and $a_p^m$'s (the superscript
$m$ is used for the eigenstate $|\mathcal{M}\rangle$) are the 
corresponding coefficients. Utilizing this relation one can determine
persistent charge currents for discrete energy levels, and therefore,
at absolute zero temperature net current carried by the ring will be
the sum of individual contributions of some specific energy levels
associated with the electron filling. This approach requires direct
evaluation of energy eigenvalues and eigenstates, like other conventional
methods available in literature~\cite{gefen,spl,maiti2,bouz,giam,yu,
maiti8,maiti10}. To avoid it, we reframe the above current expression 
(Eq.~\ref{ap4}) in terms of Green's functions introducing the concept of 
current density, instead of defining conventional currents for individual 
energy levels. The prescription is as follows.

We start with the Green's function of the ring
$\mbox{\boldmath$G$}^r = \left(\mbox{\boldmath$E$}-\mbox{\boldmath$H$} 
+i \eta \mbox{\boldmath$I$}\right)^{-1}$. It leads to
\begin{eqnarray}
\mbox{\boldmath$G$}_{ij}^r &=& \langle i|\mbox{\boldmath$G$}^r|j\rangle 
\nonumber \\
 &=& \sum_m \langle i|\mbox{\boldmath$G$}^r|\mathcal{M}\rangle \langle 
\mathcal{M}|j\rangle
\nonumber \\
 &=& \sum_m \langle i|\left(\mbox{\boldmath$E$}-\mbox{\boldmath$H$} 
+ i \eta \mbox{\boldmath$I$}\right)^{-1}|\mathcal{M}\rangle \langle 
\mathcal{M}|j\rangle
\nonumber \\
 &=& \sum_m \frac{\langle i|\mathcal{M}\rangle \langle \mathcal{M}|j\rangle}
{E-\mathcal{E}_m+i\eta} \nonumber \\
 &=& \sum_m \frac{a_i^m \, a_j^{m\,*}}{E - \mathcal{E}_m + i\eta},
\label{ap5}
\end{eqnarray}
where, $|\mathcal{M}\rangle$'s are the eigenstates of $\mbox{\boldmath$H$}$ 
satisfying the relation $\sum\limits_m |\mathcal{M}\rangle \langle 
\mathcal{M}|=\mbox{\boldmath$I$}$ and $\mathcal{E}_m$ is the eigenvalue 
for the state $|\mathcal{M}\rangle$. $a_i^m$'s
are the coefficients as described earlier. In a similar way we find,
\begin{equation}
\mbox{\boldmath$G$}_{ij}^a = \sum_m \frac{a_i^m \, a_j^{m\,*}}
{E - \mathcal{E}_m - i\eta}.
\label{ap6}
\end{equation}
Equations~\ref{ap5} and \ref{ap6} yield,
\begin{eqnarray}
\mbox{\boldmath$G$}_{ij}^r-\mbox{\boldmath$G$}_{ij}^a &=& 
\sum_m a_i^m a_j^{m\,*}\left(\frac{1}{E-\mathcal{E}_m+i\eta}
-\frac{1}{E-\mathcal{E}_m-i\eta} \right) \nonumber \\
 &=& \sum_m a_i^m a_j^{m\,*} \left(\frac{-2 i \eta}{(E-\mathcal{E}_m)^2
+\eta^2} \right) \nonumber \\
 &=& \sum_m a_i^m a_j^{m\,*} (-2 i\eta) \, \frac{\pi}{\eta} \,
\delta(E-\mathcal{E}_m) \nonumber \\
 & & \hskip 2cm(\mbox{in the limit}~ \eta \rightarrow 0^+) \nonumber \\
 &=& -2 i \pi \sum_m a_i^m a_j^{m\,*} \,\delta(E-\mathcal{E}_m).
\label{ap7}
\end{eqnarray}
Interchange of the indices $i$ and $j$ in Eq.~\ref{ap7} generates,
\begin{equation}
\mbox{\boldmath$G$}_{ji}^r-\mbox{\boldmath$G$}_{ji}^a =
-2 i \pi \sum_m a_i^{m\,*} a_j^m \,\delta(E-\mathcal{E}_m).
\label{ap8}
\end{equation}
It is clearly seen from Eqs.~\ref{ap7} and \ref{ap8} that the non-zero
contributions will only appear when the energy $E$ becomes equal to
the discrete energy eigenvalues $\mathcal{E}_m$. This immediately 
allows us to express charge current density combining Eqs.~\ref{ap4},
\ref{ap7} and \ref{ap8} as,
\begin{eqnarray}
J_c(E) &=& -\frac{et}{Nh}\sum_n \left\{\left(\mbox{\boldmath$G$}_{n+1,n}^r
-\mbox{\boldmath$G$}_{n+1,n}^a\right)e^{-i\theta} \right. \nonumber \\
 & & \left. -\left(\mbox{\boldmath$G$}_{n,n+1}^r
-\mbox{\boldmath$G$}_{n,n+1}^a\right)e^{i\theta}\right\}.
\label{ap9}
\end{eqnarray}
This is the desired expression of charge current density in terms of
Green's functions when the ring is free from SO interactions. 

\section{Green's function approach for persistent charge current density
when the ring is subjected to Rashba and Dresselhaus SO interactions} 
\label{pcso}

Next, we consider the ring with both Rashba and Dresselhaus SO interactions.
The TB Hamiltonian of the ring presented in Eq.~\ref{equ1} can be expressed
in a similar look of Eq.~\ref{ap1} like,
\begin{eqnarray}
\mbox{\boldmath$H$}_R &=& \sum_n \mbox{\boldmath$c_n^{\dagger} \epsilon_n
c_n$} + \sum_n \left(\mbox{\boldmath$c_{n+1}^{\dagger} t_{\varphi}^{n,n+1}
c_n$} e^{i\theta} \right. \nonumber \\
 & & \left. + \,\mbox{\boldmath$c_n^{\dagger} t_{\varphi}^{\dagger n,n+1}
c_{n+1}$} e^{-i\theta}\right),
\label{bp1}
\end{eqnarray}
where, different elements of the matrix 
$\mbox{\boldmath$t_{\varphi}^{n,n+1}$}$ are:
\begin{eqnarray}
\mbox{\boldmath $t_{\varphi}^{n,n+1}$}_{1,1} &=& t \nonumber \\
\mbox{\boldmath $t_{\varphi}^{n,n+1}$}_{1,2} &=& -i\,\alpha \, 
e^{-i\varphi_{n,n+1}}+\beta \, e^{i\varphi_{n,n+1}} \nonumber \\
\mbox{\boldmath $t_{\varphi}^{n,n+1}$}_{2,1} &=& -i\,\alpha \, 
e^{i\varphi_{n,n+1}}-\beta \, e^{-i\varphi_{n,n+1}} \nonumber \\
\mbox{\boldmath $t_{\varphi}^{n,n+1}$}_{2,2} &=& t. \nonumber 
\end{eqnarray}
This TB Hamiltonian leads to the charge current operator following the
prescription given in Eq.~\ref{ap2} and considering $\mbox{\boldmath$x$}=
\sum\limits_n \mbox{\boldmath$c_n^{\dagger} n c_n$}$ in the form:
\begin{eqnarray}
\mbox{\boldmath $J_{c}$} &=& \frac{2\pi i e}{N}\sum_n \left(
\mbox{\boldmath $c_n^{\dagger} t_{\varphi}^{\dagger\,n,n+1}$} 
\mbox{\boldmath $c_{n+1}$} \,e^{-i \theta}\right. \nonumber \\
 & & \left. -\, \mbox{\boldmath $c_{n+1}^{\dagger} t_{\varphi}^{n,n+1}$} 
\mbox{\boldmath $c_n$}\,e^{i \theta} \right).
\label{bp2}
\end{eqnarray}
Hence, for a particular eigenstate $|\mathcal{M}\rangle$ ($=\sum\limits_p
a_{p,\uparrow}^m |p\uparrow\rangle + a_{p,\downarrow}^m |p\downarrow\rangle$)
the charge current is written as,
\begin{eqnarray}
J_c^m &=& \frac{2\pi i e}{N} \sum_n \left(\mbox{\boldmath $a_n^{m\,*} \,
t_{\varphi}^{\dagger\,n,n+1}$} \, \mbox{\boldmath $a_{n+1}^m$} 
\,e^{-i \theta} \right. \nonumber \\
 & & \left. -\, \mbox{\boldmath $a_{n+1}^{m\,*}\, t_{\varphi}^{n,n+1}$} 
\, \mbox{\boldmath $a_n^m$}\,e^{i \theta} \right),
\label{bp3}
\end{eqnarray}
where, 
\mbox{\boldmath $a_n^m$}=$\left(\begin{array}{c}
a_{n \uparrow}^m \vspace{2mm} \\
a_{n \downarrow}^m\end{array}\right)$ and
\mbox{\boldmath $a_n^{m\,*}$}=$\left(\begin{array}{cc}
a_{n \uparrow}^{m\,*} & a_{n \downarrow}^{m\,*} 
\end{array}\right)$. 
After simplification Eq.~\ref{bp3} yields,
\begin{eqnarray}
J_c^m&=&\frac{2\pi i e}{N} \sum_n \left\{t\, a_{n,\uparrow}^{m\,*} 
a_{n+1,\uparrow}^m\, e^{-i\theta} - t\, a_{n+1,\uparrow}^{m\,*} 
a_{n,\uparrow}^m \,e^{i\theta}\right\} \nonumber \\
&+&\frac{2\pi i e}{N} \sum_n \left\{t\, a_{n,\downarrow}^{m\,*} 
a_{n+1,\downarrow}^m\,e^{-i\theta} - t\, a_{n+1,\downarrow}^{m\,*} 
a_{n,\downarrow}^m\, e^{i\theta}\right\} \nonumber \\
&+&\frac{2\pi i e}{N} \sum_n \left\{\left(i\alpha e^{-i\varphi_{n,n+1}} 
-\beta e^{i\varphi_{n,n+1}}\right) \right. \nonumber \\
 & & \left. \hskip 2cm \times \, a_{n,\uparrow}^{m\,*} a_{n+1,\downarrow}^m 
e^{-i\theta} \right. \nonumber \\
&+&\left. \left(i\alpha e^{i\varphi_{n,n+1}}+\beta
e^{-i\varphi_{n,n+1}}\right) a_{n+1,\downarrow}^{m\,*} a_{n,\uparrow}^m 
e^{i\theta}\right\} \nonumber \\
&+&\frac{2\pi i e}{N} \sum_n \left\{\left(i\alpha e^{i\varphi_{n,n+1}} 
+\beta e^{-i\varphi_{n,n+1}}\right) \right. \nonumber \\
 & & \left. \hskip 2cm \times \, a_{n,\downarrow}^{m\,*} a_{n+1,\uparrow}^m
e^{-i\theta} \right. \nonumber \\
&+&\left. \left(i\alpha e^{-i\varphi_{n,n+1}}-\beta 
e^{i\varphi_{n,n+1}}\right) a_{n+1,\uparrow}^{m\,*} a_{n,\downarrow}^m
e^{i\theta}\right\}. 
\label{bp4}
\end{eqnarray}
With this explicit expression (Eq.~\ref{bp4}) and following the above
prescription described in Appendix~\ref{pc}, we eventually get the
final result Eq.~\ref{chcurr} for persistent charge current density in 
presence of SO fields.

\section{Green's function approach for polarized spin current density
when the ring is subjected to Rashba and Dresselhaus SO interactions} 
\label{psso}

Finally, we derive the expression of persistent spin current density
in presence of SO fields. To do this we begin with the spin current 
operator,
\begin{equation}
\mbox{\boldmath $J_{s}$}=\frac{1}{2N}\left(\mbox{\boldmath ${\sigma}$}
\mbox{\boldmath ${\dot{x}}$}+ \mbox{\boldmath ${\dot{x}}$} \mbox{\boldmath
${\sigma}$}\right),
\label{cp1}
\end{equation}
where, $\mbox{\boldmath ${\sigma}$}=\{\mbox{\boldmath ${\sigma_x}$},
\mbox{\boldmath ${\sigma_y}$},\mbox{\boldmath ${\sigma_z}$}\}$. Along the 
spin quantized direction ($+Z$) this equation (Eq.~\ref{cp1}) reduces to,
\begin{equation}
\mbox{\boldmath $J_{s}^z$}=\frac{1}{2N}\left(\mbox{\boldmath ${\sigma_z}$}
\mbox{\boldmath ${\dot{x}}$}+ \mbox{\boldmath ${\dot{x}}$} \mbox{\boldmath
${\sigma_z}$}\right).
\label{cp2}
\end{equation}
Now, using the TB form given in Eq.~\ref{bp1} and following the same 
prescription of Eq.~\ref{ap1} we can express Eq.~\ref{cp2} doing a
straightforward and somewhat lengthy algebra as,
\begin{eqnarray}
\mbox{\boldmath $J_{s}^z$} &=& \frac{i\pi}{N}\sum_n \left(
\mbox{\boldmath $c_n^{\dagger}\sigma_z t_{\varphi}^{\dagger\,n,n+1}$} 
\mbox{\boldmath $c_{n+1}$} \,e^{-i \theta}\right. \nonumber \\
& - &\left. \mbox{\boldmath $c_{n+1}^{\dagger}\sigma_z t_{\varphi}^{n,n+1}$} 
\mbox{\boldmath $c_n$}\,e^{i \theta} \right) \nonumber \\
&+& \frac{i\pi}{N}\sum_n \left(\mbox{\boldmath $c_n^{\dagger} 
t_{\varphi}^{\dagger\,n,n+1} \sigma_z$} \mbox{\boldmath $c_{n+1}$} 
\,e^{-i \theta}\right. \nonumber \\
& - &\left. \mbox{\boldmath $c_{n+1}^{\dagger} t_{\varphi}^{n,n+1} \sigma_z$} 
\mbox{\boldmath $c_n$}\,e^{i \theta} \right). 
\label{cp3}
\end{eqnarray}
This spin current operator under the operation 
$\langle \mathcal{M}| \mbox{\boldmath$J_s^z$} |\mathcal{M}\rangle$ 
yields persistent spin current for a particular eigenstate 
$|\mathcal{M}\rangle$,
\begin{eqnarray}
J_s^{z,m} &=& \frac{2\pi i t}{N}\sum_n\left\{a_{n,\uparrow}^{m\,*}
a_{n+1,\uparrow}^m\,e^{-i\theta} - a_{n+1,\uparrow}^{m\,*} a_{n,\uparrow}^m 
\,e^{i\theta}\right\}\nonumber \\
&-&\frac{2\pi i t}{N} \sum_n \left\{a_{n,\downarrow}^{m\,*} 
a_{n+1,\downarrow}^m \,e^{-i\theta} - a_{n+1,\downarrow}^{m\,*} 
a_{n,\downarrow}^m \,e^{i\theta}\right\}.\nonumber \\
\label{cp4}
\end{eqnarray}
This relation leads to the final result Eq.~\ref{spcurr} following the
approach given in Appendix~\ref{pc}.


\begin{thebibliography}{99}

\bibitem{datta1} S. Datta, {\em Electronic transport in mesoscopic systems},
Cambridge University Press, Cambridge (1995).

\bibitem{imry1} Y. Imry, {\em Introduction to mesoscopic physics}, Oxford 
University Press, Oxford (1997).

\bibitem{datta2} S. Datta, {\em Quantum transport: Atom to transistor},
Cambridge University Press, Cambridge (2005).

\bibitem{yang} N. Bayers and C. N. Yang, Phys. Rev. Lett. \textbf{7},
46 (1961).

\bibitem{butt1} M. B\"{u}ttiker, Y. Imry, and R. Landauer, Phys. Lett. A
\textbf{96}, 365 (1983).

\bibitem{gefen} H. F. Cheung, Y. Gefen, E. K. Reidel, and W. H. Shih,
Phys. Rev. B \textbf{37}, 6050 (1988).

\bibitem{ambe} V. Ambegaokar and U. Eckern, Phys. Rev. Lett. \textbf{65},
381 (1990).

\bibitem{schm1} A. Schmid, Phys. Rev. Lett. \textbf{66}, 80 (1991).

\bibitem{schm2} U. Eckern and A. Schmid, Europhys. Lett. \textbf{18},
457 (1992).

\bibitem{bary} H. Bary-Soroker, O. Entin-Wohlman, and Y. Imry, Phys. 
Rev. B \textbf{82}, 144202 (2010).

\bibitem{maiti4} S. K. Maiti, J. Chowdhury and S. N. Karmakar, J. Phys.:
Condens Matter \textbf{18}, 5349 (2006).

\bibitem{maiti5} S. K. Maiti, J. Chowdhury and S. N. Karmakar, Phys.
Lett. A \textbf{332}, 497 (2004).

\bibitem{maiti6} S. K. Maiti and A. Chakrabarti, Phys. Rev. B \textbf{82},
184201 (2010).

\bibitem{maiti7} P. Dutta, S. K. Maiti, and S. N. Karmakar, Phys. Lett. A
\textbf{378}, 1388 (2014).

\bibitem{maiti9} S. K. Maiti, J. Chowdhury, and S. N. Karmakar, Solid 
State Commun. \textbf{135}, 278 (2005).

\bibitem{skm1} S. K. Maiti, Phys. Status Solidi B \textbf{248}, 1933 (2011).

\bibitem{skm2} S. K. Maiti, Int. J. Mod. Phys. B \textbf{21}, 179 (2007).

\bibitem{skm3} S. K. Maiti, Quantum Matter \textbf{3}, 413 (2014).

\bibitem{skm4} S. K. Maiti, J. Chowdhuri, and S. N. Karmakar, Synthetic
Metals \textbf{155}, 430 (2005).

\bibitem{levy} L. P. L\'{e}vy, G. Dolan, J. Dunsmuir, and H. Bouchiat, 
Phys. Rev. Lett. \textbf{64}, 2074 (1990).

\bibitem{jari} E. M. Q. Jariwala, P. Mohanty, M. B. Ketchen, and R. A.
Webb, Phys. Rev. Lett. \textbf{86}, 1594 (2001).

\bibitem{bir} N. O. Birge, Science \textbf{326}, 244 (2009).

\bibitem{chand} V. Chandrasekhar, R. A. Webb, M. J. Brady, M. B. Ketchen,
W. J. Gallagher, and A. Kleinsasser, Phys. Rev. Lett. \textbf{67}, 3578
(1991).

\bibitem{mailly1} W. Rabaud, L. Saminadayar, D. Mailly, K. Hasselbach, A.
Benoit, and B. Etienne, Phys. Rev. Lett. \textbf{86}, 3124 (2001).

\bibitem{mailly} D. Mailly, C. Chapelier, and A. Benoit, Phys. Rev. Lett.
\textbf{70}, 2020 (1993).

\bibitem{blu} H. Bluhm, N. C. Koshnick, J. A. Bert, M. E. Huber, and
K. A. Moler, Phys. Rev. Lett. \textbf{102}, 136802 (2009).

\bibitem{smt} R. A. Smith and V. Ambegaokar, Europhys. Lett. \textbf{20}, 
161 (1992).

\bibitem{mont} H. Bouchiat and G. Montambaux, J. Phys. (Paris) \textbf{60},
2695 (1989).

\bibitem{ph} E. Gambetti-C\'{e}sare, D. Weinmann, R. A. Jalabert, and 
Ph. Brune, Europhys. Lett. \textbf{60}, 120 (2002).

\bibitem{qf1} Q.-F. Sun, X. C. Xie and J. Wang, Phys. Rev. Lett. 
\textbf{98}, 196801 (2007).

\bibitem{vla} V. Vlaminck and M. Bailleul, Science \textbf{322}, 410 (2008).

\bibitem{and1} K. Ando, S. Takahashi, K. Harii, K. Sasage, J. Ieda, 
S. Maekawa, and E. Saitoh, Phys. Rev. Lett. \textbf{101}, 036601 (2008).

\bibitem{and2} K. Ando, H. Nakayama, Y. Kajiwara, D. Kikuchi, K. Sasage,
K. Uchida, K. Ikeda, and E. Saitoh, J. Appl. Phys. \textbf{105}, 07C913 
(2009).

\bibitem{rash} Y. A. Bychkov and E. I. Rashba, JETP Lett. \textbf{39},
78 (1984).

\bibitem{dress} G. Dresselhaus, Phys. Rev. \textbf{100}, 580 (1955).

\bibitem{gini} L. Meier, G. Salis, I. Shorubalko, E. Gini, S. Schon,
and K. Ensslin, Nature Phys. \textbf{4}, 77 (2008).

\bibitem{prem} J. Premper, M. Trautmann, J. Henk, and P. Bruno, Phys.
Rev. B \textbf{76}, 073310 (2007).

\bibitem{wnk} R. Winkler, {\em Spin-orbit coupling effects in
two-dimensional electron and hole systems}, Springer (2003).

\bibitem{sng} J. S. Sheng and K. Chang, Phys. Rev. B \textbf{74}, 235315
(2006).

\bibitem{spl} J. Splettstoesser, M. Governale, and U. Z\"{u}licke,
Phys. Rev. B \textbf{68}, 165341 (2003).

\bibitem{loss} D. Loss, P. Goldbart, and A. V. Balatsky, Phys. Rev. Lett. 
\textbf{65}, 1655 (1990).

\bibitem{nit} J. Nitta, T. Akazaki, H. Takayanagi, and T Enoki,
Phys. Rev. Lett. \textbf{78}, 1335 (1997). 

\bibitem{yi} C. Yin, B. Shen, Q. Zhang, F. Xu, N. Tang, L. Cen, X. Wang,
Y. Chen, and J. Yu, Appl. Phys. Lett. \textbf{97}, 181904 (2010). 

\bibitem{sc} M. Scheid, M. Kohda, Y. Kunihashi, K. Richter, and J. Nitta,
Phys. Rev. Lett. \textbf{101}, 266401 (2008).

\bibitem{stu} M. Studer, M. P. Walser, S. Baer, H. Rusterholz, 
S. Sch\"{o}n, D. Schuh, W. Wegscheider, K. Ensslin, and G. Salis,
Phys. Rev. B \textbf{82}, 235320 (2010).

\bibitem{eg} J. Schliemann, J. C. Egues, and D. Loss, Phys. Rev. Lett. 
\textbf{90}, 146801 (2003).

\bibitem{zh} B. A. Bernevig, J. Orenstein, and S.-C. Zhang, Phys. Rev. 
Lett. \textbf{97}, 236601 (2006). 

\bibitem{maiti1} S. K. Maiti, S. Sil, and A. Chakrabarti, Phys. Lett. A
\textbf{376}, 2147 (2012).

\bibitem{maiti2} S. K. Maiti, M. Dey, S. Sil, A. Chakrabarti, and S. N.
Karmakar, Europhys. Lett. \textbf{95}, 57008 (2011).

\bibitem{maiti3} S. K. Maiti, J. Appl. Phys. \textbf{110}, 064306 (2011).

\bibitem{bouz} G. Bouzerar, D. Poilblanc and G. Montambaux, Phys.
Rev. B \textbf{49}, 8258 (1994).

\bibitem{giam} T. Giamarchi and B. S. Shastry, Phys. Rev. B \textbf{51},
10915 (1995).

\bibitem{yu} N. Yu and M. Fowler, Phys. Rev. B \textbf{45}, 11795 (1992).

\bibitem{maiti8} S. K. Maiti, S. Saha, and S. N. Karmakar, Eur. Phys. 
J. B \textbf{79}, 209 (2011).

\bibitem{maiti10} S. K. Maiti, Solid State Commun. \textbf{150}, 
2212 (2010).

\bibitem{skm5} S. Saha, S. K. Maiti, and S. N. Karmakar, Eur. Phys. 
J. B \textbf{85}, 283 (2012).

\bibitem{qf2} Q.-F. Sun and X. C. Xie, Phys. Rev. B \textbf{72}, 245305 
(2005).

\bibitem{maiti11} S. Sil, S. K. Maiti, and A. Chakrabarti, Phys. Rev. Lett. 
\textbf{101}, 076803 (2008).

\bibitem{maiti12} S. Sil, S. K. Maiti, and A. Chakrabarti, Phys. Rev. B 
\textbf{78}, 113103 (2008).

\bibitem{maiti13} B. Pal, S. K. Maiti, and A. Chakrabarti, Europhys. Lett. 
\textbf{102}, 17004 (2013).

\end{thebibliography}
\end{document}